\documentclass[aps,prl,twocolumn,showpacs,groupedaddress]{revtex4-1}

\usepackage{amsfonts,amsmath,amssymb,mathrsfs}
\usepackage{hyperref}
\usepackage{subfigure}
\usepackage{color}
\usepackage{graphicx}  
\usepackage{dcolumn}   
\usepackage{bm}        
\usepackage[english]{babel}
\def\a{\alpha}

\def\r{\rho}
\def\s{\sigma}
\def\t{\tau}
\def\m{\mu}
\def\n{\nu}
\def\k{\kappa}
\def\th{\theta}
\def\g{\gamma}\def\G{\Gamma}
\def\L{\Lambda}\def\l{V}
\def\D{\Delta}
\def\la{\langle}
\def\ra{\rangle}
\def\o{\omega}\def\O{\Omega}
\def\d{\delta}
\def\p{\partial}

\def\oxthree{{\cal O}(x^3) }

\def\half{\textstyle{\frac{1}{2}}}

\def\bdoc{\begin{document}}
\def\edoc{\end{document}}
\def\bea{\begin{equation}}
\def\eea{\end{equation}}

\def\beq{\begin{eqnarray}}
\def\eeq{\end{eqnarray}}
\def\ben{\begin{enumerate}}
\def\een{\end{enumerate}}
\def\la{\langle}
\def\ra{\rangle}
\def\a{\alpha}
\def\g{\gamma}\def\G{\Gamma}
\def\d{\delta}\def\D{\Delta}
\def\e{\epsilon}
\def\z{\zeta}

\def\th{\theta}
\def\k{\kappa}
\def\l{\lambda}
\def\m{\mu}
\def\n{\nu}
\def\o{\omega}
\def\p{\pi}
\def\r{\rho}
\def\s{\sigma}
\def\t{\tau}
\def\L{{\cal L}}
\def\S{\Sigma }
\def\gsim{\; \raisebox{-.8ex}{$\stackrel{\textstyle >}{\sim}$}\;}
\def\lsim{\; \raisebox{-.8ex}{$\stackrel{\textstyle <}{\sim}$}\;}
\def\gtrsim{\gsim}
\def\lessim{\lsim}
\def\loc{{\rm local}}
\def\vm{v_{\rm max}}
\def\bh{\bar{h}}
\def\del{\partial}
\def\nab{\nabla}
\def\half{{\textstyle{\frac{1}{2}}}}
\def\fourth{{\textstyle{\frac{1}{4}}}}

\def\bD{{\bf D}}
\def\bE{{\bf E}}
\def\bF{{\bf F}}
\def\bB{{\bf B}}
\def\bP{{\bf P}}
\def\bV{{\bf v}}
\def\bv{{\bf v}}
\def\bx{{\bf x}}
\def\by{{\bf y}}
\def\bz{{\bf z}}
\def\ba{{\bf a}}
\def\bd{{\bf d}}
\def\bs{{\bf s}}
\def\bn{{\bf n}}
\def\bp{{\bf p}}

\def\O{\Omega}

\def\br{{\bf r}}
\def\bnab{{\bf \nab}}

\def\tE{\tilde{E}}
\def\tL{\tilde{L}}
\def\Horava{Ho\v{r}ava }

\def\oxtwo{\mathscr{O}\left(x^2\right)}
\def\oxthree{\mathscr{O}\left(x^3\right)}
\def\oxfour{\mathscr{O}\left(x^4\right)}
\def\oxfive{\mathscr{O}\left(x^5\right)}

\def\ph{\phantom}

\def\LL{Lanczos-Lovelock}

\begin{document}

\title{Entropy increase during physical processes for black holes in Lanczos-Lovelock gravity }
\author{Sanved Kolekar}\email{sanved@iucaa.ernet.in}\author{T. Padmanabhan}\email{paddy@iucaa.ernet.in} \affiliation{IUCAA, Pune University Campus, Ganeshkhind,\\
Pune 411007, India.} \author{Sudipta Sarkar}\email{sudiptas@imsc.res.in}
\affiliation{The Institute of Mathematical Sciences, Chennai, India}
\date{\today}

\begin{abstract}
We study quasi-stationary physical process for black holes within the context of Lanczos-Lovelock gravity. We show that the Wald entropy of the stationary black holes in Lanczos-Lovelock gravity monotonically increases for quasi-stationary physical processes in which the horizon is perturbed by the accretion of positive energy matter and the black hole ultimately settles down to a stationary state. This result
reinforces the physical interpretation of Wald entropy for \LL\ models and takes a step towards proving the analogue of the black hole area increase-theorem in a wider class of gravitational theories.
\end{abstract}
\maketitle

Pioneering work by Bekenstein \cite{Bekenstein}, Hawking \cite{Hawking}, Davies \cite{Davies} and Unruh \cite{Unruh} in the seventies showed that there is a consistent manner in which one can associate thermodynamical variables with horizons in general relativity (GR). This association gave substance to the formal connection between the laws of black hole dynamics and thermodynamics.

A natural question is  whether this analogy is a peculiar property of GR or a robust feature of any generally covariant theory of gravity. Following this line of thought Wald and collaborators \cite{Wald:1993nt,Iyer:1994ys} established the equilibrium state version of first law for black holes for any arbitrary diffeomorphism invariant theory of gravity. The black hole entropy was proposed to be the Noether charge associated with the Killing isometry generating the horizon.

The standard results in the case of general relativity concerning the entropy of horizons rely, in one way or another, on the fact that the entropy is proportional to the horizon area. This proportionality does not hold for the Wald entropy in more general theories and, therefore, it is quite intriguing that many of the results connecting gravitational dynamics to horizon thermodynamics still allows a natural generalization to a more general class of models.

Recent work suggests that this connection may indicate a far deeper truth regarding the nature of gravity viz. that it could be an emergent phenomena like, for example, fluid mechanics \cite{TProp,TPlessons}. Studies show that this correspondence, between gravitational dynamics and horizon thermodynamics, transcends general relativity and holds true for a much wider class of theories called the \LL\ models of gravity \cite{Lanczos-Lovelock:1971yv}. These are the only natural generalization of Einstein's theory to higher dimension if we insist that the equations of motion should not be of degree higher than two. The \LL\ gravity is also free from perturbative ghosts \cite{Zwiebach:1985uq} and admits consistent initial value formulation. As a result, \LL\ theories can be thought of as a natural extension of general relativity in higher dimensions. On the other hand, while \LL\ models show remarkable structural similarity with Einstein's theory, the form of the horizon entropy in the \LL\ models is quite complicated and in general entropy is not proportional to any simple geometric quantity.

Implicit in the investigations which uses the Wald entropy in these theories is the assumption that the entropy associated with a horizon behaves like ordinary thermodynamic entropy. But, the equilibrium state version of first law for black holes, established by Wald and collaborators \cite{Wald:1993nt,Iyer:1994ys} requires the existence of a stationary black hole with regular bifurcation surface. As a result, from the equilibrium state version of first law, it is not immediately clear whether the Wald entropy always increases under physical processes, except for black holes in GR, in which the ``area theorem'' asserts that area of a black hole can not decrease in any process provided null energy condition holds for the matter fields \cite{Hawking:1971vc}. The area theorem, in turn, follows  from Raychaudhuri equation and crucially depends on the contracted Einstein's equation $R_{ab} k^a k^b = 8 \pi\, T_{ab} k^a k^b$ where $k^a$ is the tangent to the horizon. Since  the entropy of black holes is no longer proportional to area in \LL\ models of gravity, there is no obvious assurance that the entropy still obeys an increase theorem. As a result, the question of validity of the second law of black hole thermodynamics for arbitrary theory of gravity remains an unresolved issue. As of yet, there is still no proof of the analog of Hawking's area theorem beyond Einstein's gravity except in the case of $f(R)$ models of gravity \cite{Jacobson:1995uq} where it was argued that the second law holds for all diffeomorphism invariant gravity theories in the quasi-stationary cases. \\

For the thermodynamic interpretation to be valid, we would expect horizon entropy to increase when a black hole in the \LL\ model participates in some  physical process, like, e.g.,  accretion of matter.
Recently, a direct proof of the physical process version of first law is proposed for Einstein-Gauss-Bonnet (EGB) gravity \cite{Chatterjee:2011wj} which establishes that the net change of black hole entropy during a physical process is positive as long as matter satisfies null energy condition. \\
In this paper, we investigate this question for general \LL\ models and show that during a physical process, the Wald entropy of stationary black holes in general Lanczos-Lovelock gravity monotonically increases provided the matter stress energy tensor obeys null energy condition. As a result, not only the net change of the entropy is positive, but the entropy is increasing at every cross section of the horizon. \textit{Our result, therefore provides a crucial step towards (possibly) proving an analogue of the area theorem in \LL\ models.}\\
Let us start with a brief review of the properties of stationary, non-extremal, Killing horizons. Let us denote the event horizon  which is a null hyper-surface as ${\cal H}$ and further take $\lambda$ to be the affine parameter. We define the vector field $k^a = (\partial_\lambda)^a$ to be the tangent to the horizon and satifying the null geodesic equation. The horizon ${\cal H}$ is foliated by $\lambda =$ constant space-like slices. Any point $p$ on these space-like hypersurfaces can then be assigned coordinates $\{\lambda, x^A\}$  where $x^A, \,(A=2, \cdots ,D)$ are the remaining transverse co-ordinates. We define $l^a$ to be the auxiliary null vector corresponding to $k^a$ satisfying the normalization $l^a k_a = -1$. These vector fields $\{k^a, l^a, e^{a}_{A}\}$ together form a basis. The induced metric is then defined as $\gamma_{ab} = g_{ab} + 2 k_{(a} l_{b)}$ satisfying $k^a \gamma_{ab}  = 0= l^a \gamma_{ab}$. The evolution of the induced metric from one slice to another is described by the following equation \cite{Wald:1984rg},
\beq
{\cal L}_{k} \gamma_{ab} = 2 \left( \sigma_{ab} + \frac{\theta}{(D-2)} \gamma_{ab} \right),\label{metric_evolution}
\eeq
where $\sigma_{ab}$ is the shear and $\theta$ is the expansion of the horizon. When the event horizon is also a Killing horizon \cite{comment1}, one can define the surface gravity $\kappa$ of the horizon through $\xi^a \nabla_a \xi^b = \kappa\, \xi^b$ where the orbits of a Killing field $\xi^a = (\partial/\partial v)^a$ are the horizon generators on the horizon. Further, the expansion and shear vanish for a stationary spacetime and using the evolution equations for $\sigma_{ab}$ and $\theta$, we obtain \cite{Wald:1984rg, Vega:2011ue},

\beq
R_{ab} k^a k^b = \xi^a \gamma^{b}_{i}\gamma^{c}_{j} \gamma^{d}_{k} R_{abcd} = k^a k^c \gamma^{b}_{m} \gamma^{d}_{n} R_{abcd} = 0. \label{stationary_conditions}
\eeq
Note that, in order to derive these relationships, we have only used the fact that the horizon is stationary Killing horizon with zero expansion and shear without any further symmetry.

We would like to consider the situation when a stationary black hole is perturbed by a weak matter stress energy tensor  and ultimately settles down to a stationary state in the asymptotic future. In such a case, the vector field $\xi^a$ becomes an exact Killing vector at late times. We assume the accretion process to be slow such that the changes in the dynamical quantities are of first order in some suitable bookkeeping parameter $\epsilon$. We further neglect all viscous effects. More specifically, we assume $\theta \sim \sigma_{ab}\sim {\cal O}(\epsilon)$.

In GR, a concrete example of such a physical process is a black hole of mass $M$ slowly accreting matter for a finite time and ultimately settle down to a stationary state. Then a linearized version of the Raychaudhuri equation gives,
\beq
\frac{d \theta}{d \lambda} \approx - R_{ab} k^a k^b = -8\, \pi \, T_{ab} k^a k^b, \label{physical_law}
\eeq
where, we have used Einstein's equation to get the second equality. If the matter stress tensor satisfies null energy condition, i. e. $ T_{ab} k^a k^b \geq 0$, the rate of change of the expansion is negative on any slice prior to the asymptotic future. Since the expansion vanishes in the future, the generators must have positive expansion during the accretion process. As a result, the area is monotonically increasing in the physical process. Note that, the result is crucially dependent on the field equation. As a result, the monotonicity of the horizon area is only valid in case of GR. Our aim is to prove a same statement for the Wald entropy during a dynamical change of the black holes in Lanczos-Lovelock gravity.

We shall now turn our attention to the features of Lanczos-Lovelock gravity. As discussed before, a natural generalization of the Einstein-Hilbert Lagrangian is provided by the Lanczos-Lovelock Lagrangian, which is the sum of dimensionally extended
Euler densities,
\beq
{\cal L}^{D} = \sum \limits_{m=0}^{[D-1)/2]} \alpha_m {\cal L}_{m}^{D},
\eeq
 where the $\alpha_m$ are arbitrary constants and ${\cal L}_{m}^{D}$ is the $m$-th order Lanczos-Lovelock term given by,
\begin{equation}
{\cal L}_{m}^{D}=\frac{1}{16\pi} \sum \limits_{m=0}^{[D-1)/2]}  \frac{1}{2^m} \delta^{a_1 b_1 \ldots a_m b_m}_{c_1 d_1 \ldots c_m d_m} R^{c_1 d_1}_{~ a_1 b_1} \cdots R^{c_m d_m}_{~ a_m b_m},
\label{actionLL}
\end{equation}
where $R^{c d}_{~ a b}$ is the $D$ dimensional curvature tensor and the generalized alternating tensor $\delta^{\ldots}_{\ldots}$ is totally anti-symmetric in both set of indices. The Einstein-Hilbert Lagrangian is a special case of Eq.~(\ref{actionLL}) when $m=1$.
The field equation of Lanczos-Lovelock theory is, $G_{ab}/(16 \pi) + \alpha_m E_{(m)ab} = (1/2) T_{ab}$ where,
\begin{eqnarray}
E^{i}_{(m) j} = - \frac{1}{16\pi} \frac{1}{2^{m+1}} \delta^{i a_1 b_1 \ldots a_m b_m}_{j c_1 d_1 \ldots c_m d_m} R^{c_1 d_1}_{~ a_1 b_1} \cdots R^{c_m d_m}_{~ a_m b_m},
\end{eqnarray}
and $m \geq 2$. For convenience,  we have written the GR part (i.e. for $m = 1$) separately so that the GR limit can be easily verified by setting all $\alpha_{m}$'s to zero.

Spherically symmetric black hole solutions in \LL\ gravity was derived in \cite{Boulware:1985wk,Boulware:1986dr} and the Wald entropy associated with a stationary Killing horizon is \cite{Visser:1993qa, Jacobson:1993xs, Visser:1993nu},
\beq
S = \frac{1}{4} \int \rho~ \sqrt{\gamma}~dA \label{entropyGB},
\eeq
where the entropy density
\beq
\rho = \left( 1 + \sum \limits_{m=2}^{[D-1)/2]} 16 \pi m \alpha_m ~^{(D-2)}L_{(m-1)} \right).
\eeq
The integration is over $(D-2)$-dimensional space-like cross-section of the horizon and $^{(D-2)}L_{(m-1)}$ is the intrinsic $(m-1)$-th Lanczos-Lovelock scalar of the horizon cross-section. We would like to prove that this entropy always increases when a black hole is perturbed by a weak matter stress energy tensor of ${\cal O}(\epsilon)$ provided the matter obeys null energy condition.

The change in entropy is \cite{Jacobson:1995uq},
\beq\label{change_wald_ent}
\Delta\,S &=&\frac{1}{4}\int_{{\cal H}} \left(\frac{d\rho}{d\lambda} +\theta\, \rho \right)\,d\lambda \,\sqrt{\gamma}\, dA.
\eeq
We define a quantity $\Theta$ as,
\beq
\Theta = \left(\frac{d\rho}{d\lambda} +\theta\, \rho \right).
\eeq
 In case of GR, $\Theta$ is equal to the expansion parameter of the null generators. But, in case of a general gravity theory, $\Theta$ is the rate of change of the entropy associated with a infinitesimal portion of horizon (see \cite{Jacobson:1995uq} for similar construction in $f(R)$ gravity). We would like to prove that given null energy condition holds, $\Theta$ is positive on any slice in a physical process.
To proceed further, we note that the change of the $(D-2)$-dimensional scalar $^{(D-2)}L_{(m-1)}$ can be thought of due to the change in the intrinsic metric. Then, we can calculate this change by using the standard result of variation of Lanczos-Lovelock scalar. The variation of $^{(D-2)}L_{(m-1)}$ simply gives the equations of motion of $(m-1)$-th order Lanczos-Lovelock term in $(D-2)$ dimensions.
Therefore, for a general Lanczos-Lovelock gravity, we can write
\beq
\frac{d\rho}{d\lambda} &=& \sum \limits_{m=2}^{[D-1)/2]} 16 \pi m \alpha_m \ k^a \nabla_a ( ^{(D-2)}L_{(m-1)})  \nonumber \\
&=& - \sum \limits_{m=2}^{[D-1)/2]} 16 \pi m \alpha_m \ ^{(D-2)}{\cal R}^{ab}_{(m-1)}\, {\cal L}_{k} \gamma_{ab}, \label{variation_section}
\eeq
\\
where we have ignored a surface term which does not contribute because the sections of the horizon are compact surfaces without boundaries. $^{(D-2)}{\cal R}_{ab}$ is the generalization of Ricci tensor for $(m-1)$-Lanczos-Lovelock gravity and is given by \cite{Kothawala:2009kc},
\beq
^{(D-2)} {\cal R}_{b \ (m-1)}^{a} &=&   \frac{1}{16 \pi} \frac{(m-1)}{2^{m}} \delta^{ a_1 b_1 \ldots a_m b_m}_{b c_1 d_1 \ldots c_m d_m} {}^{(D-2)}R_{~ a_1 b_1}^{a d_1} \cdots  \nonumber \\ ~~~~&& \cdots{}^{(D-2)} R_{~ a_m b_m}^{c_m d_m}.
\eeq
Then using Eq.~(\ref{metric_evolution}), we obtain,
\begin{eqnarray}
\Theta = \theta &+& 16 \pi \sum \limits_{m=2}^{[D-1)/2]} \alpha_m \biggl[-2\biggl(\frac{^{(D-2)}{\cal R}_{(m-1)} \theta}{(D-2)} \nonumber \\
&+& ^{(D-2)}{\cal R}^{ab}_{(m-1)} \sigma_{ab} \biggr) + \theta ~^{(D-2)}L_{(m-1)} \biggr].
\end{eqnarray}
We would like to study the rate of change of $\Theta$ along the congruence using Raychaudhuri equation and the evolution equation of shear \cite{Wald:1984rg}. Since we will be interested in keeping  quantities upto first order in perturbation over a background stationary spacetime, we perform the following operation on a product of two quantities $X$ and $Y$ to get 
\beq
X Y \approx  X^{(B)} \, Y^{(P)} +  X^{(P)} \,Y ^{(B)},\label{perturbation scheme}
\eeq
where $X^{(B)}$ is evaluated on the stationary background and $X^{(P)}$ is the value of $X$ which is linear in perturbation. We will further use the fact that for stationary background, we have $R^{(B)}_{ab} k^a k^b = 0$ using Raychaudhuri equation. Since $T^{(B)}_{ab} k^a k^b$ also vanishes, we get $E^{(B)}_{(m)ab} k^a k^b = 0$. Further, to simplify our calculation, we use the diffeomorphism freedom to make the null geodesic generators of the event horizon of the perturbed black hole coincide with those of the background stationary black hole \cite{Gao:2001ut}.

Using the perturbation scheme mentioned above and the evolution equation of $\theta$ and $\sigma_{ab}$ to linear order as $d\theta/d\lambda \approx - R^{(P)}_{ab} k^a k^b$ and $d\sigma_{ab}/d\lambda \approx  C^{(P)}_{acdb} k^c k^d$ and further using the relationships in Eq.(\ref{stationary_conditions}) for the background, the evolution equation of $\Theta$ to linear order in perturbation can be written as
\beq
\frac{d \Theta}{d\lambda} = - 8\pi\, T_{ab} k^a k^b + {\cal D}_{ab}k^a k^b,
\eeq
in which we have defined
\begin{eqnarray}
 {\cal D}_{ab}k^a k^b &=& \sum \limits_{m=2}^{[D-1)/2]}16 \pi \alpha_m \biggl[ E^{(P)}_{(m) ab} k^a k^b \nonumber \\ &+& 2m \,^{(D-2)}E^{(B)ab}_{(m-1)}\,
R^{(P)}_{acbd} k^{c}k^{d}\biggr].
\label{Ddefn}
\end{eqnarray}
Here, we have used expression of the perturbed Weyl tensor in terms of curvature and Ricci tensors and the relation ${}^{(D-2)}E_{(m)ab} = {}^{(D-2)}{\cal R}_{(m) ab} - (1/2)\gamma_{ab} {}^{(D-2)}L_{(m)}$.
We will next prove that the first order part of ${\cal D}_{ab}k^a k^b $ vanishes identically. To show this let us start with the first term in Eq.(\ref{Ddefn}) and write its first order perturbed part as:
\begin{eqnarray}
E^{(P)}_{(m) ab}k^a k^b &=& - \frac{m}{16 \pi} \frac{1}{2^{m+1}} \delta^{i a_1 b_1 \ldots a_{m-1} b_{m-1}ab}_{j c_1 d_1 \ldots c_{m-1} d_{m-1} cd} \nonumber \\ && R^{(B)c_1 d_1}_{~ a_1 b_1}
 \cdots R^{(P) cd}_{ab} k^j k_i.
\label{eperturbed}
\end{eqnarray}
Then, we first expand the background curvature tensors in the basis $\{k^a, N^a, \gamma^{a}_{b}\}$ on the horizon and use Eq.(\ref{stationary_conditions}). We also use the fact that due to the antisymmetry of the generalized alternating tensor $\delta^{\ldots}_{\ldots}$, any component of a curvature tensor along the direction of the generator of the horizon in the expression of $ E^{(P)}_{(m) ab}k^a k^b$  will not contribute. These constrains ensure that the only surviving contribution will be from the transverse components and we will finally obtain,
\begin{eqnarray}
E^{(P)}_{(m) ab}k^a k^b &=& - \frac{m}{16 \pi} \frac{1}{2^{m+1}} \delta^{i A_1 B_1 \ldots a b }_{j C_1 D_1 \ldots c d}\nonumber \\ && {}^{(D-2)}R^{(B)C_1 D_1}_{~ A_1 B_1}
\cdots R^{(P) cd}_{ab} k^j k_i,
\end{eqnarray}
where we have the fact that for stationary spacetimes \cite{Gourgoulhon:2005ng}
\begin{eqnarray}
 \gamma^m_a \gamma^n_b \gamma^c_p\gamma^d_q R^{(B) pq}_{mn} \overset{{\cal H}}{\equiv} \ ^{(D-2)}R^{(B)cd}_{ab},
\end{eqnarray}
which holds for any spacelike cross section of the stationary horizon. Next, we use the technique in \cite{Kothawala:2009kc} to write the alternating tensor in a factorized form as,
\begin{eqnarray}
 && \delta^{i A_1 B_1 \ldots A_{m-1} B_{m-1}ab}_{j C_1 D_1 \ldots C_{m-1} D_{m-1} cd}R^{(P) cd}_{ab} k^j k_i \nonumber \\
&&= - 4\, \delta^i_c  \delta^a_j \delta^{A_1 B_1 \ldots A_{m-1} B_{m-1}b}_{C_1 D_1 \ldots C_{m-1} D_{m-1} d} R^{(P) cd}_{ab} k^j k_i.
\end{eqnarray}
Using this, we finally get,
\begin{eqnarray}
 E^{(P)}_{(m) ab} k^a k^b = -2m ^{(D-2)}E^{(B)ab}_{(m-1)}\,R^{(P)}_{acbd} k^{c}k^{d}.\label{first_term}
\end{eqnarray}
Eq.~(\ref{first_term}) immediately shows that the first order part of ${\cal D}_{ab} k^a k^b$ vanishes identically and we finally arrive at,
\beq
\frac{d \Theta}{d\lambda} = - 8\pi\, T_{ab} k^a k^b + {\cal O}(\epsilon^2). \label{final_form}
\eeq
Eq.(\ref{final_form}) shows that if the null energy condition holds, the rate of change of $\Theta$ is always negative during a slow classical dynamical process (i.e. ignoring the terms which are higher order in the perturbation) which perturbs the black hole and leads to a new stationary state. Since, the final state is assumed to be stationary, both $\theta$ and $\sigma$ and as a consequence, $\Theta$ vanishes in the asymptotic future. Hence, we can use the same argument as with the expansion parameter in case of GR to conclude that $\Theta$ must be positive at every slice during the physical process. As a result, we conclude that the horizon entropy of black holes in \LL\ gravity is a monotonically increasing function during any quasi-stationary physical process, i.e.
\beq
\frac{d S}{d \lambda} \geq 0. \label{final_result}
\eeq
which is what we set out to prove.\\
In case of a dynamical scenario, it is possible to write down several candidates for the black hole entropy beyond GR \cite{Jacobson:1993vj}, such that all the expressions have same stationary limit. We have actually chosen a particular expression and the validity of Eq.(\ref{final_result}) favors such a choice. In fact, in ref.~\cite{Iyer:1994ys}, a local and geometrical prescription for the entropy of dynamical black holes is proposed. This proposal is based on a boost invariant construction and agrees with the Wald's Noether charge formula for stationary black holes and their perturbations. Interestingly, for \LL\ gravity, the entropy expression used in this work matches with the expression obtained from the boost invariant construction. Consequently, our result provides a strong justification in favor of the prescription for dynamical entropy as proposed in ref.~\cite{Iyer:1994ys}. This may also be important  to decide the right candidate for the entropy of non stationary black holes for non \LL\ gravity models.

From a quantum gravity perspective, one naturally associates the black hole entropy to the micro-states of the black hole. In any reasonable theory of gravity with stable black hole solutions, the density of states are expected to be of the form $\exp{(S_{W})}$, where $S_{W}$ is the corresponding Wald entropy. As an example, consider the case of string theory where it has been demonstrated at least for extremal and near-extremal black holes, that the microscopic computations lead to expressions exactly matching with the corresponding Wald's formula for entropy \cite{Sen:2007qy, Lopes Cardoso:1998wt}. Hence, it is quite desirable that the Wald entropy satisfies an increase theorem.

Some obvious further investigations suggested by this work are the
following: First, one would like to relax the quasi-stationarity
physical process assumption and calculate the full change of the
Wald entropy along the horizon to understand the validity of
classical second law for Lanczos-Lovelock models. The possible
conclusions depend crucially on the signs of the higher order
terms in Eq.(\ref{final_form}). As in case of GR, if all the
higher order terms are negative, this would imply that $\Theta$
has to decrease monotonically. Further, it cannot be negative on
any cross section of the horizon which otherwise will lead to
$\Theta \to -\infty$ and hence the existence of a caustic on the
event horizon which is prohibited by the fact that the event
horizon is future complete \cite{penrose}. Then, to avoid the
contradiction, $\Theta$ must be positive on any arbitrary slice of
the horizon which would lead to classical second law for \LL\
gravity.

 The second issue worth exploring is the following: A
crucial assumption in our derivation is that there exists a
quasi-stationary physical process in which the black hole
ultimately settle down to a final stationary state. Although such
an assumption is quite reasonable, one should not overlook an
extreme possibility that the black holes in a general \LL\ gravity
may not be stable under a small perturbation. While the linear
stability around flat spacetime of the Einstein-Gauss-Bonnet
gravity is demonstrated in \cite{Deser:1989jm}, the positive
energy theorem has not been extended to a general Lovelock theory
and even if such an extension is possible, there may be other
instabilities. This requires further investigation.

Finally, we would like to note that the techniques used in this
work are specific to \LL\ gravity. As a result, it would be
worthwhile to find a general approach which can answer whether
classical second law holds in a physical process for \textit{any}
diffeomorphism invariant gravity theory or applies to a special
class of action functionals.

\section*{Acknowledgments}
SK is supported by a Fellowship from the Council of Scientific and Industrial Research (CSIR), India. The authors would like to thank Ted Jacobson for useful comments and suggestions. SS would also like to thank G Date and A Chatterjee for discussions. SS thanks IUCAA for hospitality during his visit when this work was completed. TP's research is partially supported by J.C.Bose Research grant.

\end{document}